\documentclass[reprint, superscriptaddress]{revtex4-1}

\usepackage{graphicx}
\usepackage{tabularx}

\begin{document}

\title{Formation of Strain-Induced Quantum Dots in Gated Semiconductor Nanostructures}
\author{Ted Thorbeck}
\affiliation{Quantum Measurement Division, NIST, Gaithersburg, Maryland, USA}
\affiliation{Joint Quantum Institute and Dept. of Physics, University of Maryland, College Park, Maryland, USA}
\author{Neil M. Zimmerman}
\affiliation{Quantum Measurement Division, NIST, Gaithersburg, Maryland, USA}

\begin{abstract}
Elastic strain changes the energies of the conduction band in a semiconductor, which will affect transport through a semiconductor nanostructure. We show that the typical strains in a semiconductor nanostructure from metal gates are large enough to create strain-induced quantum dots (QDs). We
simulate a commonly used QD device architecture, metal gates on bulk silicon, and show the formation of strain-induced QDs.  The strain-induced QD can be eliminated by replacing the metal gates with poly-silicon gates. Thus strain can be as important as electrostatics to QD device operation operation.
\end{abstract}

\maketitle

Because elastic strain changes the band structure of a crystal, it is deliberately used in many silicon nanostructures\cite{yu_fundamentals_1999}. For example, a silicon superlattice can be made with periodic strains rather than a heterostructure\cite{euaruksakul_influence_2008}; this has the advantage of avoiding materials interfaces. Another example is in phosphorus donors in silicon, which are being studied as qubits\cite{kane_silicon-based_1998, pla_single-atom_2012}. Strain changes the hyperfine coupling\cite{huebl_phosphorus_2006}, so local strains can be used to address individual qubits\cite{dreher_electroelastic_2011}. As a final example, strain increases the mobility of electrons in silicon\cite{sun_physics_2007, niquet_effects_2012}, which has led to strain engineering in the channel of modern silicon transistors. Each of these three examples uses intentional strains to alter the band structure of silicon to enable or improve a silicon nanostructure.  In this paper we will consider the unintentional effects of strain arising from metal gates and contacts in a semiconductor nanostructure when operating a device at low temperatures. Specifically, we will show that the strain-altered conduction band (CB) can explain previously observed but unexplained quantum dots (QDs). This means that the effect on the CB from the strain from a gate or contact can be as important as the electrostatic effect of the gate or contact.

Electrical transport through QDs, which are nanometer-scale regions confined in all three dimensions, are enabling exciting physics. A lot of work is being done to make silicon QDs to build an electron-spin based quantum computer\cite{morton_embracing_2011, zwanenburg_silicon_2012}, because the spin of an electron in a silicon QD has a long coherence time\cite{maune_coherent_2012, pla_single-atom_2012}.  Silicon QDs are also being pursued for electrical standards because charge pumps built from silicon QDs are more stable as a function of time than metallic charge pumps\cite{zimmerman_electrical_2003, zimmerman_why_2008}.

To pursue these applications, many different methods are used to create silicon QDs\cite{zwanenburg_silicon_2012}. In this letter we will focus on one common method of creating silicon QDs: metallic gates on bulk silicon. In this method voltages are applied to a patterned set of metallic gates to form tunnel barriers and QDs in the silicon below\cite{maune_coherent_2012, nordberg_enhancement-mode_2009, hu_electron_2009, angus_gate-defined_2007, shaji_spin_2008, xiao_measurement_2010}.  This method of creating QDs is attractive because the QDs are tunable and the device architecture is similar to modern silicon transistors. A major problem for this method of creating QDs is that it is common to observe many-electron QDs where there should be a tunnel barrier\cite{nordberg_enhancement-mode_2009, hu_electron_2009, angus_gate-defined_2007, thorbeck_determining_2012}. Previously, these QDs have been attributed to charged defects such as dopants and interface traps\cite{nordberg_enhancement-mode_2009, hu_electron_2009, angus_gate-defined_2007}.  Although steps have been taken to reduce the interface trap and dopant density, this has not eliminated these QDs.  Furthermore, when the Fermi level is near the CB the charged defects are acceptor-like (negatively charged). Thus the defect will prevent electrons from localizing nearby \cite{Schroder_Semiconductor_2006}. Also, these QDs are routinely observed in the same location in different devices, which is inconsistent with QDs caused by randomly located charge traps.  We can explain why many-electron QDs are observed in the same location in different devices by considering strain from the metal gates.

In this letter, we will show that metal gates, which are routinely used to electrostatically create QDs in silicon nanostructures, will also create strains large enough to induce a QD. First, we will go through a general argument that suggests that the typical strain from putting metals on a semiconductor nanostructure can be large enough to create strain-induced QDs. This general argument applies to many different materials systems and architectures. Then, we will simulate a device with metal gates on bulk silicon to show that the strains can induce a QD.  The location of the strain-induced QD can explain why QDs are frequently observed where there should be only an electrostatic tunnel barrier. The strain-induced QD should either be harnessed or elminated.   We will discuss the potential advantages of strain-induced QDs. Then we will show how the strain-induced QD can be eliminated by replacing the metal with highly-doped poly-silicon. To demonstrate that strain-induced QDs can be a problem for many different architectures, we consider another device architecture in the supplementary information: silicon nanowires with metal contacts.

A lot of work has been done previously in optical strain-induced QDs in III-Vs by placing small stressors on top of a quantum well\cite{boxberg_theory_2007, lipsanen_optical_2004}. Strain from oxidizing a mesa-etched nanowire has been shown to cause tunnel barriers\cite{ono_fabrication_2002, horiguchi_mechanism_2001}, at the ends of the nanowire. Strain from lattice mismatch is needed to explain the properties of resonant tunneling diodes in Si/SiGe nanowire heterostructures\cite{akyuz_inhomogeneous_1998}. It has been suggested that strains from lattice defects can be problematic in Si/SiGe QDs\cite{evans_nanoscale_2012}. In contrast, we focus on the impact of the strain from metal gates, which are routinely used without consideration of the elastic strain caused by them.

We will first go through a general argument which could apply to many different semiconductor nanostructures. This will also establish the theoretical framework used in the later simulated examples.  We begin by discussing the physical origin and typical magnitude of the strain. Then, we will discuss how strain changes the energy of the CB. Finally, we will discuss when the change in the CB minimum is enough to induce QDs.

Strain is inevitable in a semiconductor QD device. The strain may arise from how the device was manufactured or from operating at cryogenic temperature. During fabrication, for example, growing a thermal oxide on a silicon nanowire will induce stress in the nanowire, because thermally grown SiO$_2$ must expand to incorporate the extra oxygen atoms. Most of this volume expansion occurs perpendicular to the growth plane, but some remains as compressive strain in the SiO$_2$\cite{kobeda_intrinsic_1987}. Strain can also come from cooling the device to its cryogenic operating temperature. Because no silicon QD consists only of silicon, the silicon must have interfaces. Coefficient of thermal expansion (CTE) mismatch at the interface will cause strain when cooled. For example, the CTEs of aluminum and silicon are 23 x 10$^{-6}$ K$^{-1}$ and 2.6 x 10$^{-6}$ K$^{-1}$. This mismatch can setup strains as large as 0.6 $\%$ for a 300 K change in temperature. 

Figure 1 shows schematically how CTE mismatch can strain a device. In this example, metal is deposited on top of a semiconductor. Metals typically have larger CTEs than semiconductors, so the metal will contract more than the semiconductor when cooled. Far away from the semiconductor-metal interface, the metal is free to contract, but near the interface the semiconductor prevents the metal from contracting, causing tensile elastic strain in the metal. Conversely, the semiconductor side of the interface is under compressive elastic strain. 

The CB minimum will change linearly with strain\cite{fischetti_band_1996}.  For electrons in the $ \pm k_z $ valleys (see supplementary information), the change in energy of the CB minimum as a function of strain is
\begin{equation}
\Delta E_C  = \Xi_u \epsilon_z + \Xi_d ( \epsilon_x + \epsilon_y + \epsilon_z )
\end{equation}	
where $ \epsilon_x $ , $ \epsilon_y $ and $ \epsilon_z $  are the components of the strain, the relative change in length, and the deformation potentials are $ \Xi_u$ = 10.5 eV  and $ \Xi_d$ = 1.1 eV\cite{fischetti_band_1996}. Because $ \Xi_u >> \Xi_d $ , the first term dominates this equation, thus $ \Delta E_C \approx  \Xi_u \epsilon_x $. Thus, the energies of the $ \pm k_z $ valleys will change by about 10 meV for every 0.1 \% strain in $ \epsilon_z $. The change in energies of the $ \pm k_x $ and  $ \pm k_y $ valleys can be determined by replacing $ \epsilon_z $ in the first term of the right-hand-side with $ \epsilon_x $ and $ \epsilon_y$. To explain the physical origin of the deformation potential, we first need to understand which atomic orbitals make up the CB. Near the valleys the CB has a significant contribution from the bonding 3d-orbitals.  Because these are bonding levels, the energy of the bond decreases  $ (\Delta E_C < 0) $ as the atoms are brought closer together $ (\epsilon < 0) $.  Therefore, the deformation potentials of the CB are positive.

\begin{figure}
\includegraphics{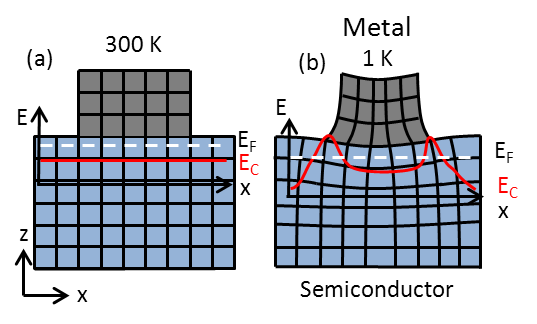}
\caption{\label{fig:figure1}Thermal contraction of metal on top of a semiconductor from (a) room temperature to (b) cryogenic temperatures. We also show a schematic of the change of the CB ($E_C$) in the silicon immediately below the metal, showing peaks near the corners of the metal box due to stress concentration.}
\end{figure}

Now that we know the typical magnitude of the strain and understanding how strain modifies the CB, we will return to our simple example. Figure 1(b) shows a schematic of the effect of the strains on the CB. Strains in the semiconductor near the metal raise the CB with respect to the CB far away from the metal, where there is no elastic strain.  The peaks in the CB beneath the corners of the metal are due to stress concentration at the corners of the metal.  There is a local minimum of the CB between the peaks because there is less strain underneath the center of the metal. 

To determine if this strain-altered CB can induce a QD, we need to consider both the shape and the magnitude of the confining potential. The shape of the CB in figure 1(b) can result in a QD, because electrons can be trapped in the local minimum (underneath the center of the metal), and the peaks (underneath the corners of the metal) form tunnel junctions. To trap an electron, the confining potential (barrier height) must be larger than kT and the charging energy (the amount of energy it takes to add an electron on the QD). At cryogenic temperatures kT is less than 0.1 meV. A typical charging energy for a QD of this size is   ̴ $ \sim $ 1 meV.  Therefore, 10 meV, the typical magnitude of the strain-induced change in the CB is large enough to confine electrons. In fact, a barrier height of 10 meV is the same magnitude as an electrostatic tunnel barrier\cite{sellier_subthreshold_2007} and the change in CB due to interface traps\cite{ nordberg_enhancement-mode_2009}. This example shows that, for a wide range of semiconductor nanostructures, CTE mismatch can lead to strains that have the right shape and magnitude to induce a QD.  In the supplementary information we consider the resistance of the tunnel barriers (of order M$\Omega$) and discuss the robustness of the barriers with respect to gate voltage changes (robust to changes of order 0.1 V).

Figure 2(a) shows a device architecture for electrons in a surface-gated bulk-silicon device. This architecture consists of a bulk silicon (lightly p-doped) wafer covered in 10 nm of thermally grown SiO$_2$, with two aluminum gates (upper gate, UG, and lower gate, LG) on top\cite{maune_coherent_2012, nordberg_enhancement-mode_2009, hu_electron_2009, angus_gate-defined_2007, shaji_spin_2008, xiao_measurement_2010}. The two aluminum gates are perpendicular to each other and are isolated from each other by 3 nm of AlO$_x$. The UG is 80 nm tall and 50 nm wide, and the LG has a 25 nm diameter. A positive voltage on the UG will cause an inversion layer a few nanometers thick to form at the Si-SiO$_2$ interface. Current flows through the inversion layer (current flows from and to a heavily doped source and drain regions that are far from the LG).    A negative voltage on LG can deplete the silicon below the LG to form a single tunnel barrier, directly below the LG. However, QDs are commonly observed in this location, where there should only be a tunnel barrier\cite{maune_coherent_2012, nordberg_enhancement-mode_2009, hu_electron_2009, angus_gate-defined_2007, shaji_spin_2008, xiao_measurement_2010}. In this section, we will show how strain from the CTE mismatch can induce a QD directly below the LG. 

\begin{figure}
\centering
\includegraphics{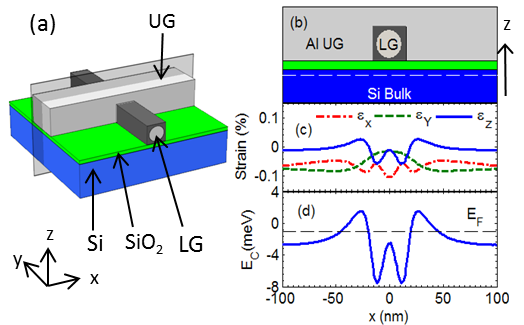}
\caption{\label{fig:figure3}Calculated strains and CB change for metal gates on bulk silicon. (a) A pseudo-3D drawing of the device showing the bulk  silicon wafer (blue), SiO$_2$ (green), Al UG and LG (grey) and AlO$_x$ (dark grey), not to scale. (b) Cross section of the device through the translucent plane in (a) with same colors as (a), not to scale vertically. Dashed white line represents the inversion layer (1 nm below the Si-SiO$_2$ interface), electrons flow in the inversion layer from left to right. (c) Calculated strains in inversion layer (white line in (b)), showing the effect of strains from CTE mismatch of Al and AlO$_x$. (d) CB modulation from strains in (c) showing tunnel barriers at $x = \pm$ 30 and a QD between. The dashed line represents the Fermi level. (b)-(d) all have the same horizontal axis. }
\end{figure}

Figure 2(b) shows the simulated strains for this architecture. We use COMSOL multiphysics \cite{Note1} to simulate the strains in the device (details in the supplementary information). The simulation includes both CTE mismatch and intrinsic stress in the SiO$_2$ (-200 MPa\cite{kobeda_intrinsic_1987}). Unlike figure 1, where the CTE mismatch between metal and semiconductor caused the strain, here CTE mismatch between Al and AlO$_x$ creates stresses that propagate into the silicon below.  Al (23x10$^{-6}$ K$^{-1}$) has a much larger CTE than AlO$_x$ (5.4x10$^{-6}$ K$^{-1}$). For a change in temperature from 293 K to 1 K, Al is in tensile stress because the AlO$_x$ is preventing it from contracting. Conversely, the Al is putting the AlO$_x$ in compressive stress. These stresses propagate through the SiO$_2$ into the silicon. CTE mismatch from the SiO$_2$ and intrinsic stress from the SiO$_2$ only results in uniform strain and so cannot induce a QD.

Confining electrons in an inversion layer breaks the six-fold valley degeneracy and only the $\pm k_z$ valleys are occupied (supplementary information). In Fig. 2(d) we use eq. 1 to calculate the change in energy of the $\pm k_z$ valleys due to the strains shown in Fig. 2(c). (Because the first term dominates eq. 1, $\Delta E_C$ has the same shape as $\epsilon_z$) The peaks at $ x = \pm $ 30 nm form tunnel barriers for electrons. The height (4 meV) and length (40 nm) of these barriers give them tunneling resistance of 20 M$\Omega$ (See supplementary information). A strain-induced QD forms in the dip between these barriers.  This QD is directly below the LG, which can explain the previously observed QDs\cite{maune_coherent_2012, nordberg_enhancement-mode_2009, hu_electron_2009, angus_gate-defined_2007, shaji_spin_2008, xiao_measurement_2010}. 

We have shown that in this device architecture strain can induce a QD. Because strain is often ignored when designing the device, the strain-induced QD would show up as additional QDs.  This could explain why such unintentional QDs are a common problem in surface-gated architectures\cite{nordberg_enhancement-mode_2009, hu_electron_2009, angus_gate-defined_2007, thorbeck_determining_2012}. Strain-induced QDs have several advantages over electrostatically defined QDs. Making electrostatic QDs requires additional metal gates, limiting the number of QDs that can be operated. Each gate also makes the QD bigger, which makes it harder to reach the few-electron limit. These advantages can be obtained in an architecture which is already being used to make electrostatic QDs, when taking into account strain effects. 

\begin{figure}
   \begin{center}
   \includegraphics[width=\columnwidth]{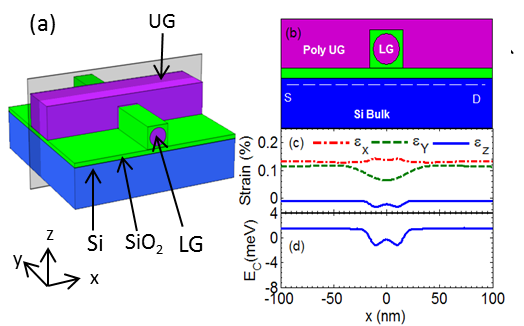}
   \end{center}
\caption{\label{fig:figs2}Calculated strains and CB change for poly-Si gates on bulk Si. (a) A pseudo-3D image of the device showing the bulk silicon wafer (blue), SiO$_2$ (green), Poly-Si LG and UG (purple), not to scale. (b) Cross section of the device through the translucent plane in (a) with same colors as (a), not to scale vertically. Dashed white line represents the inversion layer. (c) Calculated strains in inversion layer (1 nm below the Si-SiO$_2$ interface). (d) CB modulation from strains in (c). Note that the change in the CB is much smaller than in Fig. 2(d) due to changing the gate material from aluminum to poly-silicon. (b)-(d) all have the same horizontal axis. }
\end{figure}

If the strain-induced QD is not desirable, then it should be eliminated. In the geometry of Fig. 2, the strain-induced QD can be eliminated by replacing the Al and AlO$_X$ in the gate stack with heavily doped poly-silicon and SiO$_2$. Electrostatically this device operates just like the Al gated device. This material switch, from Al (CTE 23 x 10$^{-6}$ K$^{-1}$) to poly-Si (2.9 x 10$^{-6}$ K$^{-1}$), reduces the CTE mismatch by an order of magnitude. Figure 3 shows the strains calculated for the same geometry and dimensions as for the Al gate device. The strains due to the LG are much smaller than in the Al gated device. This results in a much smaller modulation of the CB due to strain (Fig. 3(d)). Because the peaks that had been the tunnel junctions in the Al gated device are now only 0.1 meV high, this device will not form a strain-induced QD.

We have shown that strain from CTE mismatch can cause QDs.  Although we only showed one example geometry, our argument is more general. Most metals have a larger CTE than insulators or semiconductors. Thus our qualitative argument that the typical strains in a nanostructure at low temperatures are large enough to create strain-induced QDs, applies to other material systems such as carbon, germanium and III-Vs.  To demonstrate this, in the supplementary information we consider another example, a silicon nanowire with metal contacts. Because the typical strains in a nanostructure can induce QDs, we suggest that the strains should either be used or ameliorated. Reducing the strain (perhaps by replacing metal gates with poly-Si gates) would eliminate the strain-induced QDs, allowing an electrostatically gated device to operate as intended. In addition to avoiding unintentional effects due to strain, we can also see some advantages to strain-induced QDs compared to other methods of creating QDs. Strain-induced QDs require fewer gates, allowing them to be smaller than electrostatic QDs. 

Because strain can be as important as electrostatics to the operation of a QD device, the effects of strain should be considered when analyzing the results from or designing semiconductor nanostructures.

We would like to acknowledge helpful conversations with Michael Stewart, Panu Koppinen, Josh Pomeroy, Justin Perron and Vladimir Aksyuk. This work was supported in part by the Laboratory for Physical Sciences (EAO93195).

\bibliographystyle{aipnum}

\def\url#1{}

\cleardoublepage

\begin{center}
     \textbf{Supplementary Information}
 \end{center}

\setcounter{equation}{0} 
\setcounter{figure}{0} 

\renewcommand{\thesection}{S.\arabic{section}}

\makeatletter
\makeatletter \renewcommand{\fnum@figure}
{\figurename~S\thefigure}
\makeatother

\title{Supplemntary Information for Formation of Strain-Induced Quantum Dots in Gated Semiconductor Nanostructures}
\author{Ted Thorbeck}
\affiliation{Quantum Measurement Division, NIST, Gaithersburg, Maryland, USA}
\affiliation{Joint Quantum Institute and Dept. of Physics, University of Maryland, College Park, Maryland, USA}
\author{Neil M. Zimmerman}
\affiliation{Quantum Measurement Division, NIST, Gaithersburg, Maryland, USA}

\date{4 September 2013}

\maketitle

\subsection{Silicon band structure}

The bottom of the conduction band (CB) and top of the valence band (VB) are shown schematically in figure S1. In bulk silicon the bottom of the CB is 6-fold degenerate, while the top of the VB is 4-fold degenerate. However, these degeneracies can be broken by confinement and strain\cite{yu_fundamentals_1999, ando_electronic_1982}.

The six valleys in the CB are located at  $ \vec{k} = (2 \pi / a_0 )(\pm 0.85,0,0), (2 \pi / a_0 )(0, \pm 0.85, 0), $ and $  (2 \pi / a_0 )(0, 0, \pm 0.85) $, where $ a_0 $ is the silicon lattice constant ($ a_0 $ = 0.543 nm). Around each valley the effective mass is anisotropic. For the $ \pm k_z $ valleys, centered at $ k_0 = (2 \pi / a_0 )(0, 0, \pm 0.85) $, the effective mass in the z-direction is $ m_l $ = 0.98 $ m_e$,  while the effective mass in the x and y directions is $m_t$ = 0.19$m_e$, where $m_e$ is the free electron mass. Thus the dispersion relation for the $ \pm k_z $ valleys is
\begin{equation}
\Delta E_C(\vec{k})  = \frac{\hbar^2}{2}( \frac{k_x^2}{m_t}+\frac{k_y^2}{m_t}+\frac{(k_z - k_0)^2}{m_l} )
\end{equation}	
The dispersion relations for the $ \pm k_x $ and $ \pm k_y $ valleys are similar. 

To show how confinement breaks the six-fold valley degeneracy [Fig. S1(b)], consider an inversion layer perpendicular to the z-axis. We can neglect lateral confinement because of the much larger length scale. The potential well confining the electrons can be approximated by an infinite triangular well\cite{ando_electronic_1982}. In the effective mass approximation, the eigenenergies for the electron are given by
\begin{equation}
E_n=(\frac{(e\mathcal{E})^2 \hbar^2}{2m_z})^{\frac{1}{3}} (-a_n)		
\end{equation}
where $e\mathcal{E}$ is the slope of the quantum well ($\mathcal{E}$ is the electric field confining the electron), and $a_n$ are the zeroes of the Airy function ($a_0 \approx$ -2.33). For the $\pm k_z$ valleys, $m_z = m_l$, while the other four valleys ($ \pm x$, $\pm y$) $m_z = m_t$. Therefore, the $\pm k_z$ valleys have a lower energy ($E_0$ = 37 meV for $\mathcal{E}$ = 105 V/cm) than the $\pm k_x$ and $\pm k_y$ valleys ($E_0$ = 63 meV). Because this energy difference is many $kT$ at cryogenic temperatures (for $T$ = 1 K, $kT$ = 86 $\mu$eV), the $\pm k_x$ and $\pm k_y$ valleys will not be occupied.

\begin{figure}
\includegraphics{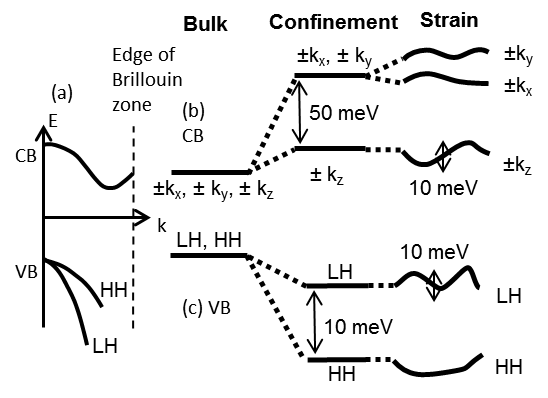}
\caption{\label{fig:figS1}Diagram of the band structure of silicon. (a) The CB has 6 equivalent valleys, and the VB has LH and HH states. (b) The 6-fold degeneracy of the CB is split by confinement, the energies of the lowest two valley states are further modulated by strain. (c) Similar diagram for VB}
\end{figure}

Strain also changes the energies of the valleys [Fig. S1(b)]. This is due to the atomic lattice being squeezed or pulled by the strain, thus raising or lowering each band within the silicon. The change in energy of the $\pm k_z$ valleys is described by the deformation potentials $\Xi_u$ = 10.5 eV and $\Xi_d$ = 1.1 eV.
\begin{equation}
\Delta E_C= \Xi_u \epsilon_z + \Xi_u (\epsilon_x + \epsilon_y + \epsilon_z)		
\end{equation}
The deformation potential, $\Xi_u$, for the CB in silicon is positive because there is a large contribution to the CB from the atomic 3-d bonding orbitals. We already showed that in an inversion layer electrons will only occupy the $\pm k_z$ valleys, so we will not further consider the $\pm k_x$ and $\pm k_y$  valleys. The local strains, set up by CTE mismatch, are typically of order 0.1 \%. Combined with the deformation potentials, these strains will change the energy of the valley by $\approx$ 10 meV. Because the CB is defined with respect to electron energy, a peak in the CB due to strain will cause a tunnel barrier for electrons.

The VB of silicon consists of a light hole (LH), heavy hole (HH), and spin orbit split off (SO) bands[1]. The SO band is 44 meV lower than the LH and HH bands. Because this splitting is many $kT$ at cryogenic temperatures, holes will not occupy the SO band, and it will not be considered further.

The dispersion relations for holes in the LH and HH are
\begin{equation}
\Delta E_V=-Ak^2\mp(B k^4 + C(k_x^2 k_y^2+k_y^2 k_z^2 + k_z^2 k_x^2 ))	
\end{equation}

where the -- is for the HH and the + for the LH and $A=- 4.25(\hbar^2 / 2m_e ), B=- 0.63(\hbar^2 / 2m_e )$ and $C=4.9(\hbar^2 / 2m_e )$\cite{yu_fundamentals_1999}. We note that $\Delta E_V \approx (3/2) a_v (\epsilon_y+ \epsilon_z )$ for the LH. The $ a_v $ term is positive in silicon because the VB consists of the atomic 3p-bonding orbitals\cite{yu_fundamentals_1999}. Confinement will split the LH and the HH [Fig. S1(c)]. However, because the VB dispersion relation is more complicated than CB dispersion relation, showing the effect of confinement is more complicated for the VB than for the CB. Restricting ourselves to the case of confinement in a nanowire, recent theoretical work\cite{csontos_spin-3/2_2009} has shown the highest VB state is predominately LH in character with the spin aligned with the axis of the nanowire. Therefore, we assume that we only need to consider the LH in our analysis.

Strain will further change the energy of the VB [Fig. S1(c)]. Ignoring the effect of band mixing due to strain, the change in the VB of silicon due to strain is 
\begin{equation}
\Delta E_V= a_v (\epsilon_x + \epsilon_y \epsilon_z) + b_v (\epsilon_x -  (\epsilon_y + \epsilon_z) / 2)		
\end{equation}
where $a_v$ = 2.1 eV and $b_v$ = -2.33 eV for silicon\cite{yu_fundamentals_1999}.  Because the VB is defined with respect to electron energy, a dip in the VB due to strain will cause a barrier for holes.

\subsection{Barrier resistance}

To observe a QD the barrier resistance, R, must be larger than the resistance quantum ($R \gg R_K = 26 k\Omega$) to observe discrete charging events on the QD\cite{grabert_single_1992}. But the resistance must not be so large as to make the current too small to measure, $R < 1 G\Omega$. In the main text we determined that the typical modulation of the CB due to strain is of order 10 meV. To calculate the tunneling resistance, we will need the length scale over which the CB changes. 10 nm is a typical gate width or oxide thickness. (The gate width is typically limited by electron beam lithography to  >  10 nm, and oxide thicknesses are typically of order 10 nm to prevent leakage.) Using a WKB (Wentzel –- Kramers –- Brillouin) tunneling rate for a parabolic barrier to determine the tunneling resistance, 
\begin{equation}
\frac{1}{R}=G=N \frac{e^2}{\hbar} e^{-\frac{\pi}{\sqrt{2}}  \frac{\sqrt{m^*}}{\hbar} \sqrt{\phi} L}		
\end{equation}
where $m^*=0.19m_0$ and taking the number of channels, N = 1 (because $N \approx wk_f \approx $1\cite{beenakker_quantum_1991}). The height and length of the barrier, $\phi$ and $L$, are conservatively assumed to be 5 meV and 20 nm. This gives us an estimated tunneling resistance of 4 M$\Omega$. This satisfies our criteria on the tunneling resistance, and thus strain-induced CB modulation can cause tunnel barriers and QDs.

\subsection{Simulation Details}
The strain was simulated using COMSOL multiphysics.  A few simplifying assumptions were made in the simulation: the silicon was simulated isotropically and room temperature materials properties were used, including Young’s modulus, Poisson’s ratio, density and CTE. All simulations were performed in three dimensions. The simulated volume was about 5x10$^7$ nm$^3$. Despite the nanoscale features atomistic simulations are not necessary; similar simulations of the stress are performed in similarly scaled commercial transistors\cite{sun_physics_2007}. A zero displacement boundary condition was used for the bottom surface of the simulated volume, and a zero force boundary condition was used everywhere else on the surface.  We verified that the boundary conditions do not affect the area of the simulation of interest by changing the size of the simulation, and observing that the strains in the area of interest were not affected.  Physically, the simulation corresponds to cooling the device adiabatically.

  \begin{tabular}{ | c | c | c | c | c | }
    \hline
    Material & Young's  & Poisson's  & Density & CTE\\ 
    & Modulus (GPa) & Ratio &  (kg/m$^3$) &  (x10$^{-6}$/K) \\ \hline 
    Si & 130 & 0.27 & 2300 & 2.6 \\ \hline
    SiO$_2$ & 73 & 0.17 & 2200 & 0.49 \\ \hline
    Al & 70 & 0.35 & 2700 & 23 \\ \hline
    AlO$_x$ & 300 & 0.22 & 3900 & 5.4 \\ \hline
    Ni & 220 & 0.31 & 8900 & 13 \\ \hline
    Poly-Si & 170 & 0.22 & 2300 & 2.9 \\ \hline
  \end{tabular}

\subsection{Silicon Nanowire with Metal Contacts}

VB holes can be confined in a chemically-grown nanowire between metallic contacts\cite{lu_semiconductor_2006, zwanenburg_spin_2009, zhong_coherent_2005, salfi_electronic_2010}. Chemically-grown nanowires are attractive because it is easy to grow a small diameter nanowire with low surface roughness\cite{lu_semiconductor_2006}. Tunnel barriers near the nanowire-contact interface confine the holes within the nanowire. These tunnel barriers are essential to form the QDs, and so understanding the reason for their existence is important. Sometimes these barriers are due to Schottky barriers; however, the metal-semiconductor pair is often deliberately chosen to prevent a Schottky barrier. For example, in bulk metal-InAs contacts the Fermi level is pinned above the CB\cite{bhargava_fermi-level_1997, scheffler_tunable_2008}, and thus a Schottky barrier should not form. Also, many other metal-semiconductor combinations, which form Schottky barriers in bulk, should not form Schottky barriers in a nanostructure, because there are not enough interface states on a nanowire to pin the Fermi level\cite{leonard_electrical_2011, leonard_size-dependent_2006}. Nevertheless, it is common to observe tunnel barriers in metal-nanowire contacts that should not have Schottky barriers\cite{dahl-nissen_comparison_2012}. This mystery has driven us to consider strain-induced tunnel barriers as an explanation.

Figure S2(a) shows a device architecture for a chemically grown nanowire that is frequently used to form QDs for holes\cite{lu_semiconductor_2006, zwanenburg_spin_2009, zhong_coherent_2005, salfi_electronic_2010}. A typical device (Fig. 2(a)) consists of an undoped Si nanowire (5 nm radius) on top of a thick SiO$_2$ layer. Contacts for the source and drain are formed with Nickel (50 nm thick separated by 200 nm). We assume that no Schottky barrier forms at the metal-nanowire interface, so holes can flow freely from the Ni into the nanowire. In this section, we will show that strain can cause tunnel barriers for holes in the nanowire near the metal contacts.

\begin{figure}
\centering
   \includegraphics[width=\columnwidth]{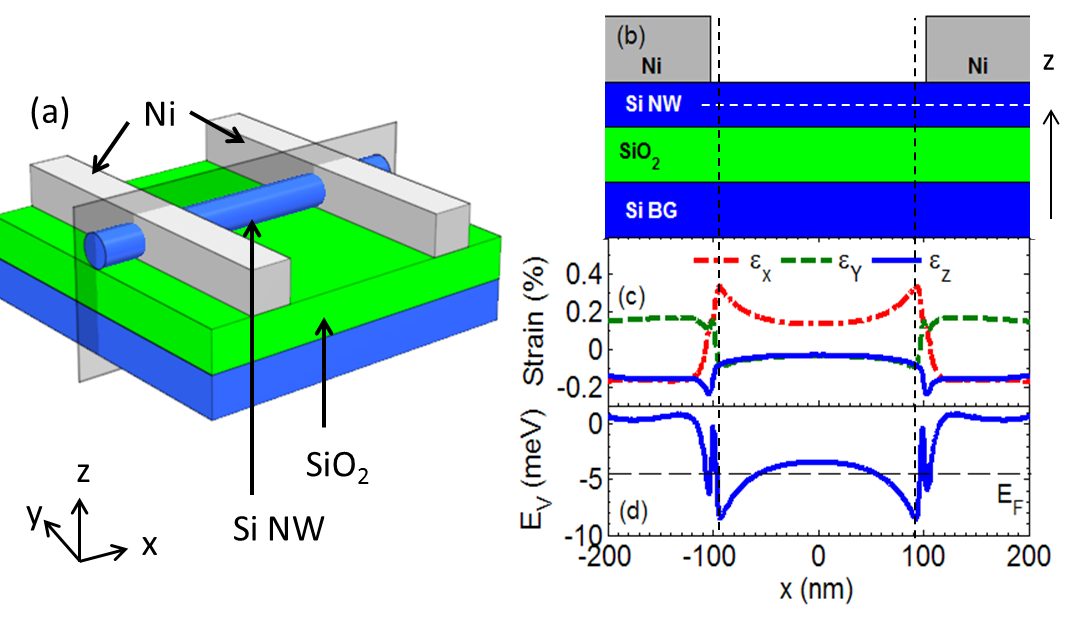}
\caption{\label{fig:figure2}Calculated strains and VB change for a chemically grown silicon nanowire with nickel contacts. (a) A pseudo-3D drawing of the device showing the bulk Si wafer (blue), SiO$_2$ (green), Si-nanowire (blue), Ni source and drain contacts (grey). (b) Cross section of the device through the translucent plane in (a) with same colors as (a), not to scale vertically. (c) Calculated strains at center of the nanowire (white dashed line in (b)), showing the effect of strains CTE mismatch of Si-nanowire and Ni contact. (d) VB modulation for LH from strains in (c) showing tunnel barriers at $x =\pm $ 100 and a QD in between. Horizontal dashed line represents the Fermi level. (b)-(d) all have the same horizontal axis. Black vertical dashed line shows the location of the tunnel junctions in (b) to (d).}
\end{figure}

Most of the strain in this device comes from the CTE mismatch of nickel (13x10$^{-6}$ $ K^{-1} $ ) and silicon (2.6x10$^{-6}$ $ K^{-1} $). Figure S2(c) shows the strains in the center of the nanowire as simulated for a change in temperature of 293 K to 1 K. Strains elsewhere in the nanowire are similar. Because the metal contacts shrink, this pulls on the nanowire. Therefore, the nanowire is stretched (in tension) in the x direction (and has compressive strains in $ \epsilon_x $ and $ \epsilon_y$ because of Poisson’s ratio).

The VB of bulk silicon consists of degenerate HH and LH bands. Confinement in a nanowire will split the LH and HH bands, and the topmost VB state has been predicted to be predominantly LH in character, with the spin quantized along the direction of the nanowire\cite{csontos_spin-3/2_2009}. We take the simulated strains and use eq. S5 to calculate the change in the VB for LHs [Fig. S2(d)]. As mentioned below eq. S5 the change in the VB is due primarily to the sum of the strains, $ \epsilon_y+ \epsilon_z$. Dips in the VB, like those at $x = \pm$ 90 nm in Fig. S2(c), which are between the edges of the metal contacts, create barriers for holes. Between the barriers, a strain-induced QD forms in the peak in the VB centered at $x$ = 0. The tunnel barriers have a height of 5 meV and a length of 30 nm which gives an estimated tunneling resistance of 45 M$\Omega$). This resistance is large enough to quantize the charge on the QD without shutting current off. So we have shown that strain can induce a QD in a nanowire between two metal contacts. We propose this as an explanation of why tunnel barriers are sometimes observed at metal-nanowire interfaces that should not form Schottky barriers\cite{dahl-nissen_comparison_2012}.

\bibliographystyle{aipnum}

\begin{thebibliography}{39}
\expandafter\ifx\csname natexlab\endcsname\relax\def\natexlab#1{#1}\fi
\expandafter\ifx\csname bibnamefont\endcsname\relax
  \def\bibnamefont#1{#1}\fi
\expandafter\ifx\csname bibfnamefont\endcsname\relax
  \def\bibfnamefont#1{#1}\fi
\expandafter\ifx\csname citenamefont\endcsname\relax
  \def\citenamefont#1{#1}\fi
\expandafter\ifx\csname url\endcsname\relax
  \def\url#1{\texttt{#1}}\fi
\expandafter\ifx\csname urlprefix\endcsname\relax\def\urlprefix{URL }\fi
\providecommand{\bibinfo}[2]{#2}
\providecommand{\eprint}[2][]{\url{#2}}

\bibitem[{\citenamefont{Yu and Cardona}(1999)}]{yu_fundamentals_1999}
\bibinfo{author}{\bibfnamefont{P.~Y.} \bibnamefont{Yu}} \bibnamefont{and}
  \bibinfo{author}{\bibfnamefont{M.}~\bibnamefont{Cardona}},
  \emph{\bibinfo{title}{Fundamentals of Semiconductors: Physics and Materials
  Properties}} (\bibinfo{publisher}{Springer-Verlag Telos},
  \bibinfo{year}{1999}), \bibinfo{edition}{2nd} ed., ISBN
  \bibinfo{isbn}{{354065352X}}.

\bibitem[{\citenamefont{Euaruksakul et~al.}(2008)\citenamefont{Euaruksakul, Li,
  Zheng, Himpsel, Ritz, Tanto, Savage, Liu, and
  Lagally}}]{euaruksakul_influence_2008}
\bibinfo{author}{\bibfnamefont{C.}~\bibnamefont{Euaruksakul}},
  \bibinfo{author}{\bibfnamefont{Z.~W.} \bibnamefont{Li}},
  \bibinfo{author}{\bibfnamefont{F.}~\bibnamefont{Zheng}},
  \bibinfo{author}{\bibfnamefont{F.~J.} \bibnamefont{Himpsel}},
  \bibinfo{author}{\bibfnamefont{C.~S.} \bibnamefont{Ritz}},
  \bibinfo{author}{\bibfnamefont{B.}~\bibnamefont{Tanto}},
  \bibinfo{author}{\bibfnamefont{D.~E.} \bibnamefont{Savage}},
  \bibinfo{author}{\bibfnamefont{X.~S.} \bibnamefont{Liu}}, \bibnamefont{and}
  \bibinfo{author}{\bibfnamefont{M.~G.} \bibnamefont{Lagally}},
  \bibinfo{journal}{Physical Review Letters} \textbf{\bibinfo{volume}{101}},
  \bibinfo{pages}{147403} (\bibinfo{year}{2008}),
  \urlprefix\url{http://link.aps.org/doi/10.1103/PhysRevLett.101.147403}.

\bibitem[{\citenamefont{Kane}(1998)}]{kane_silicon-based_1998}
\bibinfo{author}{\bibfnamefont{B.~E.} \bibnamefont{Kane}},
  \bibinfo{journal}{Nature} \textbf{\bibinfo{volume}{393}},
  \bibinfo{pages}{133} (\bibinfo{year}{1998}), ISSN \bibinfo{issn}{0028-0836},
  \urlprefix\url{http://www.nature.com/nature/journal/v393/n6681/abs/393133a0.html}.

\bibitem[{\citenamefont{Pla et~al.}(2012)\citenamefont{Pla, Tan, Dehollain,
  Lim, Morton, Jamieson, Dzurak, and Morello}}]{pla_single-atom_2012}
\bibinfo{author}{\bibfnamefont{J.~J.} \bibnamefont{Pla}},
  \bibinfo{author}{\bibfnamefont{K.~Y.} \bibnamefont{Tan}},
  \bibinfo{author}{\bibfnamefont{J.~P.} \bibnamefont{Dehollain}},
  \bibinfo{author}{\bibfnamefont{W.~H.} \bibnamefont{Lim}},
  \bibinfo{author}{\bibfnamefont{J.~J.~L.} \bibnamefont{Morton}},
  \bibinfo{author}{\bibfnamefont{D.~N.} \bibnamefont{Jamieson}},
  \bibinfo{author}{\bibfnamefont{A.~S.} \bibnamefont{Dzurak}},
  \bibnamefont{and} \bibinfo{author}{\bibfnamefont{A.}~\bibnamefont{Morello}},
  \bibinfo{journal}{Nature} \textbf{\bibinfo{volume}{489}},
  \bibinfo{pages}{541} (\bibinfo{year}{2012}), ISSN \bibinfo{issn}{0028-0836},
  \urlprefix\url{http://www.nature.com/nature/journal/v489/n7417/full/nature11449.html}.

\bibitem[{\citenamefont{Huebl et~al.}(2006)\citenamefont{Huebl, Stegner,
  Stutzmann, Brandt, Vogg, Bensch, Rauls, and
  Gerstmann}}]{huebl_phosphorus_2006}
\bibinfo{author}{\bibfnamefont{H.}~\bibnamefont{Huebl}},
  \bibinfo{author}{\bibfnamefont{A.~R.} \bibnamefont{Stegner}},
  \bibinfo{author}{\bibfnamefont{M.}~\bibnamefont{Stutzmann}},
  \bibinfo{author}{\bibfnamefont{M.~S.} \bibnamefont{Brandt}},
  \bibinfo{author}{\bibfnamefont{G.}~\bibnamefont{Vogg}},
  \bibinfo{author}{\bibfnamefont{F.}~\bibnamefont{Bensch}},
  \bibinfo{author}{\bibfnamefont{E.}~\bibnamefont{Rauls}}, \bibnamefont{and}
  \bibinfo{author}{\bibfnamefont{U.}~\bibnamefont{Gerstmann}},
  \bibinfo{journal}{Physical Review Letters} \textbf{\bibinfo{volume}{97}},
  \bibinfo{pages}{166402} (\bibinfo{year}{2006}),
  \urlprefix\url{http://link.aps.org/doi/10.1103/PhysRevLett.97.166402}.

\bibitem[{\citenamefont{Dreher et~al.}(2011)\citenamefont{Dreher, Hilker,
  Brandlmaier, Goennenwein, Huebl, Stutzmann, and
  Brandt}}]{dreher_electroelastic_2011}
\bibinfo{author}{\bibfnamefont{L.}~\bibnamefont{Dreher}},
  \bibinfo{author}{\bibfnamefont{T.~A.} \bibnamefont{Hilker}},
  \bibinfo{author}{\bibfnamefont{A.}~\bibnamefont{Brandlmaier}},
  \bibinfo{author}{\bibfnamefont{S.~T.~B.} \bibnamefont{Goennenwein}},
  \bibinfo{author}{\bibfnamefont{H.}~\bibnamefont{Huebl}},
  \bibinfo{author}{\bibfnamefont{M.}~\bibnamefont{Stutzmann}},
  \bibnamefont{and} \bibinfo{author}{\bibfnamefont{M.~S.}
  \bibnamefont{Brandt}}, \bibinfo{journal}{Physical Review Letters}
  \textbf{\bibinfo{volume}{106}}, \bibinfo{pages}{037601}
  (\bibinfo{year}{2011}),
  \urlprefix\url{http://link.aps.org/doi/10.1103/PhysRevLett.106.037601}.

\bibitem[{\citenamefont{Sun et~al.}(2007)\citenamefont{Sun, Thompson, and
  Nishida}}]{sun_physics_2007}
\bibinfo{author}{\bibfnamefont{Y.}~\bibnamefont{Sun}},
  \bibinfo{author}{\bibfnamefont{S.~E.} \bibnamefont{Thompson}},
  \bibnamefont{and} \bibinfo{author}{\bibfnamefont{T.}~\bibnamefont{Nishida}},
  \bibinfo{journal}{Journal of Applied Physics} \textbf{\bibinfo{volume}{101}},
  \bibinfo{pages}{104503} (\bibinfo{year}{2007}), ISSN
  \bibinfo{issn}{00218979},
  \urlprefix\url{http://jap.aip.org/resource/1/japiau/v101/i10/p104503_s1}.

\bibitem[{\citenamefont{Niquet et~al.}(2012)\citenamefont{Niquet, Delerue, and
  Krzeminski}}]{niquet_effects_2012}
\bibinfo{author}{\bibfnamefont{Y.-M.} \bibnamefont{Niquet}},
  \bibinfo{author}{\bibfnamefont{C.}~\bibnamefont{Delerue}}, \bibnamefont{and}
  \bibinfo{author}{\bibfnamefont{C.}~\bibnamefont{Krzeminski}},
  \bibinfo{journal}{Nano Letters} \textbf{\bibinfo{volume}{12}},
  \bibinfo{pages}{3545} (\bibinfo{year}{2012}), ISSN \bibinfo{issn}{1530-6984},
  \urlprefix\url{http://dx.doi.org/10.1021/nl3010995}.

\bibitem[{\citenamefont{Morton et~al.}(2011)\citenamefont{Morton, {McCamey},
  Eriksson, and Lyon}}]{morton_embracing_2011}
\bibinfo{author}{\bibfnamefont{J.~J.~L.} \bibnamefont{Morton}},
  \bibinfo{author}{\bibfnamefont{D.~R.} \bibnamefont{{McCamey}}},
  \bibinfo{author}{\bibfnamefont{M.~A.} \bibnamefont{Eriksson}},
  \bibnamefont{and} \bibinfo{author}{\bibfnamefont{S.~A.} \bibnamefont{Lyon}},
  \bibinfo{journal}{Nature} \textbf{\bibinfo{volume}{479}},
  \bibinfo{pages}{345} (\bibinfo{year}{2011}), ISSN \bibinfo{issn}{0028-0836},
  \urlprefix\url{http://www.nature.com/nature/journal/v479/n7373/full/nature10681.html}.

\bibitem[{\citenamefont{Zwanenburg et~al.}(2012)\citenamefont{Zwanenburg,
  Dzurak, Morello, Simmons, Hollenberg, Klimeck, Rogge, Coppersmith, and
  Eriksson}}]{zwanenburg_silicon_2012}
\bibinfo{author}{\bibfnamefont{F.~A.} \bibnamefont{Zwanenburg}},
  \bibinfo{author}{\bibfnamefont{A.~S.} \bibnamefont{Dzurak}},
  \bibinfo{author}{\bibfnamefont{A.}~\bibnamefont{Morello}},
  \bibinfo{author}{\bibfnamefont{M.~Y.} \bibnamefont{Simmons}},
  \bibinfo{author}{\bibfnamefont{L.~C.~L.} \bibnamefont{Hollenberg}},
  \bibinfo{author}{\bibfnamefont{G.}~\bibnamefont{Klimeck}},
  \bibinfo{author}{\bibfnamefont{S.}~\bibnamefont{Rogge}},
  \bibinfo{author}{\bibfnamefont{S.~N.} \bibnamefont{Coppersmith}},
  \bibnamefont{and} \bibinfo{author}{\bibfnamefont{M.~A.}
  \bibnamefont{Eriksson}}, \bibinfo{journal}{{arXiv:1206.5202}}
  (\bibinfo{year}{2012}), \urlprefix\url{http://arxiv.org/abs/1206.5202}.

\bibitem[{\citenamefont{Maune et~al.}(2012)\citenamefont{Maune, Borselli,
  Huang, Ladd, Deelman, Holabird, Kiselev, Alvarado-Rodriguez, Ross, Schmitz
  et~al.}}]{maune_coherent_2012}
\bibinfo{author}{\bibfnamefont{B.~M.} \bibnamefont{Maune}},
  \bibinfo{author}{\bibfnamefont{M.~G.} \bibnamefont{Borselli}},
  \bibinfo{author}{\bibfnamefont{B.}~\bibnamefont{Huang}},
  \bibinfo{author}{\bibfnamefont{T.~D.} \bibnamefont{Ladd}},
  \bibinfo{author}{\bibfnamefont{P.~W.} \bibnamefont{Deelman}},
  \bibinfo{author}{\bibfnamefont{K.~S.} \bibnamefont{Holabird}},
  \bibinfo{author}{\bibfnamefont{A.~A.} \bibnamefont{Kiselev}},
  \bibinfo{author}{\bibfnamefont{I.}~\bibnamefont{Alvarado-Rodriguez}},
  \bibinfo{author}{\bibfnamefont{R.~S.} \bibnamefont{Ross}},
  \bibinfo{author}{\bibfnamefont{A.~E.} \bibnamefont{Schmitz}},
  \bibnamefont{et~al.}, \bibinfo{journal}{Nature}
  \textbf{\bibinfo{volume}{481}}, \bibinfo{pages}{344} (\bibinfo{year}{2012}),
  ISSN \bibinfo{issn}{0028-0836},
  \urlprefix\url{http://www.nature.com/nature/journal/v481/n7381/full/nature10707.html?..}

\bibitem[{\citenamefont{Zimmerman and
  Keller}(2003)}]{zimmerman_electrical_2003}
\bibinfo{author}{\bibfnamefont{N.~M.} \bibnamefont{Zimmerman}}
  \bibnamefont{and} \bibinfo{author}{\bibfnamefont{M.~W.}
  \bibnamefont{Keller}}, \bibinfo{journal}{Measurement Science and Technology}
  \textbf{\bibinfo{volume}{14}}, \bibinfo{pages}{1237} (\bibinfo{year}{2003}),
  ISSN \bibinfo{issn}{0957-0233, 1361-6501},
  \urlprefix\url{http://iopscience.iop.org/0957-0233/14/8/307}.

\bibitem[{\citenamefont{Zimmerman et~al.}(2008)\citenamefont{Zimmerman, Huber,
  Simonds, Hourdakis, Fujiwara, Ono, Takahashi, Inokawa, Furlan, and
  Keller}}]{zimmerman_why_2008}
\bibinfo{author}{\bibfnamefont{N.~M.} \bibnamefont{Zimmerman}},
  \bibinfo{author}{\bibfnamefont{W.~H.} \bibnamefont{Huber}},
  \bibinfo{author}{\bibfnamefont{B.}~\bibnamefont{Simonds}},
  \bibinfo{author}{\bibfnamefont{E.}~\bibnamefont{Hourdakis}},
  \bibinfo{author}{\bibfnamefont{A.}~\bibnamefont{Fujiwara}},
  \bibinfo{author}{\bibfnamefont{Y.}~\bibnamefont{Ono}},
  \bibinfo{author}{\bibfnamefont{Y.}~\bibnamefont{Takahashi}},
  \bibinfo{author}{\bibfnamefont{H.}~\bibnamefont{Inokawa}},
  \bibinfo{author}{\bibfnamefont{M.}~\bibnamefont{Furlan}}, \bibnamefont{and}
  \bibinfo{author}{\bibfnamefont{M.~W.} \bibnamefont{Keller}},
  \bibinfo{journal}{Journal of Applied Physics} \textbf{\bibinfo{volume}{104}},
  \bibinfo{pages}{033710} (\bibinfo{year}{2008}), ISSN
  \bibinfo{issn}{00218979},
  \urlprefix\url{http://jap.aip.org/resource/1/japiau/v104/i3/p033710_s1}.



\bibitem[{\citenamefont{Nordberg et~al.}(2009)\citenamefont{Nordberg, Eyck,
  Stalford, Muller, Young, Eng, Tracy, Childs, Wendt, Grubbs
  et~al.}}]{nordberg_enhancement-mode_2009}
\bibinfo{author}{\bibfnamefont{E.~P.} \bibnamefont{Nordberg}},
  \bibinfo{author}{\bibfnamefont{G.~A.~T.} \bibnamefont{Eyck}},
  \bibinfo{author}{\bibfnamefont{H.~L.} \bibnamefont{Stalford}},
  \bibinfo{author}{\bibfnamefont{R.~P.} \bibnamefont{Muller}},
  \bibinfo{author}{\bibfnamefont{R.~W.} \bibnamefont{Young}},
  \bibinfo{author}{\bibfnamefont{K.}~\bibnamefont{Eng}},
  \bibinfo{author}{\bibfnamefont{L.~A.} \bibnamefont{Tracy}},
  \bibinfo{author}{\bibfnamefont{K.~D.} \bibnamefont{Childs}},
  \bibinfo{author}{\bibfnamefont{J.~R.} \bibnamefont{Wendt}},
  \bibinfo{author}{\bibfnamefont{R.~K.} \bibnamefont{Grubbs}},
  \bibnamefont{et~al.}, \bibinfo{journal}{Physical Review B}
  \textbf{\bibinfo{volume}{80}}, \bibinfo{pages}{115331}
  (\bibinfo{year}{2009}),
  \urlprefix\url{http://link.aps.org/doi/10.1103/PhysRevB.80.115331}.

\bibitem[{\citenamefont{Hu and Yang}(2009)}]{hu_electron_2009}
\bibinfo{author}{\bibfnamefont{B.}~\bibnamefont{Hu}} \bibnamefont{and}
  \bibinfo{author}{\bibfnamefont{C.~H.} \bibnamefont{Yang}},
  \bibinfo{journal}{Physical Review B} \textbf{\bibinfo{volume}{80}},
  \bibinfo{pages}{075310} (\bibinfo{year}{2009}),
  \urlprefix\url{http://link.aps.org/doi/10.1103/PhysRevB.80.075310}.

\bibitem[{\citenamefont{Angus et~al.}(2007)\citenamefont{Angus, Ferguson,
  Dzurak, and Clark}}]{angus_gate-defined_2007}
\bibinfo{author}{\bibfnamefont{S.~J.} \bibnamefont{Angus}},
  \bibinfo{author}{\bibfnamefont{A.~J.} \bibnamefont{Ferguson}},
  \bibinfo{author}{\bibfnamefont{A.~S.} \bibnamefont{Dzurak}},
  \bibnamefont{and} \bibinfo{author}{\bibfnamefont{R.~G.} \bibnamefont{Clark}},
  \bibinfo{journal}{Nano Letters} \textbf{\bibinfo{volume}{7}},
  \bibinfo{pages}{2051} (\bibinfo{year}{2007}), ISSN \bibinfo{issn}{1530-6984},
  \urlprefix\url{http://dx.doi.org/10.1021/nl070949k}.

\bibitem[{\citenamefont{Shaji et~al.}(2008)\citenamefont{Shaji, Simmons,
  Thalakulam, Klein, Qin, Luo, Savage, Lagally, Rimberg, Joynt
  et~al.}}]{shaji_spin_2008}
\bibinfo{author}{\bibfnamefont{N.}~\bibnamefont{Shaji}},
  \bibinfo{author}{\bibfnamefont{C.~B.} \bibnamefont{Simmons}},
  \bibinfo{author}{\bibfnamefont{M.}~\bibnamefont{Thalakulam}},
  \bibinfo{author}{\bibfnamefont{L.~J.} \bibnamefont{Klein}},
  \bibinfo{author}{\bibfnamefont{H.}~\bibnamefont{Qin}},
  \bibinfo{author}{\bibfnamefont{H.}~\bibnamefont{Luo}},
  \bibinfo{author}{\bibfnamefont{D.~E.} \bibnamefont{Savage}},
  \bibinfo{author}{\bibfnamefont{M.~G.} \bibnamefont{Lagally}},
  \bibinfo{author}{\bibfnamefont{A.~J.} \bibnamefont{Rimberg}},
  \bibinfo{author}{\bibfnamefont{R.}~\bibnamefont{Joynt}},
  \bibnamefont{et~al.}, \bibinfo{journal}{Nature Physics}
  \textbf{\bibinfo{volume}{4}}, \bibinfo{pages}{540} (\bibinfo{year}{2008}),
  ISSN \bibinfo{issn}{1745-2473},
  \urlprefix\url{http://www.nature.com/nphys/journal/v4/n7/full/nphys988.html}.

\bibitem[{\citenamefont{Xiao et~al.}(2010)\citenamefont{Xiao, House, and
  Jiang}}]{xiao_measurement_2010}
\bibinfo{author}{\bibfnamefont{M.}~\bibnamefont{Xiao}},
  \bibinfo{author}{\bibfnamefont{M.~G.} \bibnamefont{House}}, \bibnamefont{and}
  \bibinfo{author}{\bibfnamefont{H.~W.} \bibnamefont{Jiang}},
  \bibinfo{journal}{Physical Review Letters} \textbf{\bibinfo{volume}{104}},
  \bibinfo{pages}{096801} (\bibinfo{year}{2010}),
  \urlprefix\url{http://link.aps.org/doi/10.1103/PhysRevLett.104.096801}.

\bibitem[{\citenamefont{Thorbeck and
  Zimmerman}(2012)}]{thorbeck_determining_2012}
\bibinfo{author}{\bibfnamefont{T.}~\bibnamefont{Thorbeck}} \bibnamefont{and}
  \bibinfo{author}{\bibfnamefont{N.~M.} \bibnamefont{Zimmerman}},
  \bibinfo{journal}{Journal of Applied Physics} \textbf{\bibinfo{volume}{111}},
  \bibinfo{pages}{064309} (\bibinfo{year}{2012}), ISSN
  \bibinfo{issn}{00218979},
  \urlprefix\url{http://jap.aip.org/resource/1/japiau/v111/i6/p064309_s1}.

\bibitem[{\citenamefont{Schroder, D}(2006)}]{Schroder_Semiconductor_2006}
 \bibinfo{author}{\bibfnamefont{D.}~\bibnamefont{Schroder}}
  \emph{\bibinfo{booktitle}{Semiconductor Material and Device Characterization}}
  (\bibinfo{publisher}{John Wiley and Sons}, \bibinfo{year}{2006}).



\bibitem[{\citenamefont{Boxberg and Tulkki}(2007)}]{boxberg_theory_2007}
\bibinfo{author}{\bibfnamefont{F.}~\bibnamefont{Boxberg}} \bibnamefont{and}
  \bibinfo{author}{\bibfnamefont{J.}~\bibnamefont{Tulkki}},
  \bibinfo{journal}{Reports on Progress in Physics}
  \textbf{\bibinfo{volume}{70}}, \bibinfo{pages}{1425} (\bibinfo{year}{2007}),
  ISSN \bibinfo{issn}{0034-4885, 1361-6633},
  \urlprefix\url{http://iopscience.iop.org/0034-4885/70/8/R04}.

\bibitem[{\citenamefont{Lipsanen and Sopanen}(2004)}]{lipsanen_optical_2004}
\bibinfo{author}{\bibfnamefont{H.}~\bibnamefont{Lipsanen}} \bibnamefont{and}
  \bibinfo{author}{\bibfnamefont{M.}~\bibnamefont{Sopanen}}, in
  \emph{\bibinfo{booktitle}{Optics of Quantum Dots and Wires}}, edited by
  \bibinfo{editor}{\bibfnamefont{G.~W.} \bibnamefont{Bryant}} \bibnamefont{and}
  \bibinfo{editor}{\bibfnamefont{G.~S.} \bibnamefont{Solomon}}
  (\bibinfo{publisher}{Artech House}, \bibinfo{year}{2004}), p.
  \bibinfo{pages}{547}, ISBN \bibinfo{isbn}{1580537618}.

\bibitem[{\citenamefont{Ono et~al.}(2002)\citenamefont{Ono, Yamazaki, Nagase,
  Horiguchi, Shiraishi, and Takahashi}}]{ono_fabrication_2002}
\bibinfo{author}{\bibfnamefont{Y.}~\bibnamefont{Ono}},
  \bibinfo{author}{\bibfnamefont{K.}~\bibnamefont{Yamazaki}},
  \bibinfo{author}{\bibfnamefont{M.}~\bibnamefont{Nagase}},
  \bibinfo{author}{\bibfnamefont{S.}~\bibnamefont{Horiguchi}},
  \bibinfo{author}{\bibfnamefont{K.}~\bibnamefont{Shiraishi}},
  \bibnamefont{and}
  \bibinfo{author}{\bibfnamefont{Y.}~\bibnamefont{Takahashi}},
  \bibinfo{journal}{Solid-State Electronics} \textbf{\bibinfo{volume}{46}},
  \bibinfo{pages}{1723} (\bibinfo{year}{2002}), ISSN \bibinfo{issn}{0038-1101},
  \urlprefix\url{http://www.sciencedirect.com/science/article/pii/S0038110102001417}.

\bibitem[{\citenamefont{Horiguchi et~al.}(2001)\citenamefont{Horiguchi, Nagase,
  Shiraishi, Kageshima, Takahashi, and Murase}}]{horiguchi_mechanism_2001}
\bibinfo{author}{\bibfnamefont{S.}~\bibnamefont{Horiguchi}},
  \bibinfo{author}{\bibfnamefont{M.}~\bibnamefont{Nagase}},
  \bibinfo{author}{\bibfnamefont{K.}~\bibnamefont{Shiraishi}},
  \bibinfo{author}{\bibfnamefont{H.}~\bibnamefont{Kageshima}},
  \bibinfo{author}{\bibfnamefont{Y.}~\bibnamefont{Takahashi}},
  \bibnamefont{and} \bibinfo{author}{\bibfnamefont{K.}~\bibnamefont{Murase}},
  \bibinfo{journal}{Japanese Journal of Applied Physics}
  \textbf{\bibinfo{volume}{40}}, \bibinfo{pages}{L29} (\bibinfo{year}{2001}).

\bibitem[{\citenamefont{Akyüz et~al.}(1998)\citenamefont{Akyüz, Zaslavsky,
  Freund, Syphers, and Sedgwick}}]{akyuz_inhomogeneous_1998}
\bibinfo{author}{\bibfnamefont{C.~D.} \bibnamefont{Akyüz}},
  \bibinfo{author}{\bibfnamefont{A.}~\bibnamefont{Zaslavsky}},
  \bibinfo{author}{\bibfnamefont{L.~B.} \bibnamefont{Freund}},
  \bibinfo{author}{\bibfnamefont{D.~A.} \bibnamefont{Syphers}},
  \bibnamefont{and} \bibinfo{author}{\bibfnamefont{T.~O.}
  \bibnamefont{Sedgwick}}, \bibinfo{journal}{Applied Physics Letters}
  \textbf{\bibinfo{volume}{72}}, \bibinfo{pages}{1739} (\bibinfo{year}{1998}),
  ISSN \bibinfo{issn}{00036951}.

\bibitem[{\citenamefont{Evans et~al.}(2012)\citenamefont{Evans, Savage, Prance,
  Simmons, Lagally, Coppersmith, Eriksson, and
  Schülli}}]{evans_nanoscale_2012}
\bibinfo{author}{\bibfnamefont{P.~G.} \bibnamefont{Evans}},
  \bibinfo{author}{\bibfnamefont{D.~E.} \bibnamefont{Savage}},
  \bibinfo{author}{\bibfnamefont{J.~R.} \bibnamefont{Prance}},
  \bibinfo{author}{\bibfnamefont{C.~B.} \bibnamefont{Simmons}},
  \bibinfo{author}{\bibfnamefont{M.~G.} \bibnamefont{Lagally}},
  \bibinfo{author}{\bibfnamefont{S.~N.} \bibnamefont{Coppersmith}},
  \bibinfo{author}{\bibfnamefont{M.~A.} \bibnamefont{Eriksson}},
  \bibnamefont{and} \bibinfo{author}{\bibfnamefont{T.~U.}
  \bibnamefont{Schülli}}, \bibinfo{journal}{Advanced Materials}
  \textbf{\bibinfo{volume}{24}}, \bibinfo{pages}{5217–5221}
  (\bibinfo{year}{2012}), ISSN \bibinfo{issn}{1521-4095}.

\bibitem[{\citenamefont{Kobeda and Irene}(1987)}]{kobeda_intrinsic_1987}
\bibinfo{author}{\bibfnamefont{E.}~\bibnamefont{Kobeda}} \bibnamefont{and}
  \bibinfo{author}{\bibfnamefont{E.~A.} \bibnamefont{Irene}},
  \bibinfo{journal}{Journal of Vacuum Science \& Technology B: Microelectronics
  and Nanometer Structures} \textbf{\bibinfo{volume}{5}}, \bibinfo{pages}{15}
  (\bibinfo{year}{1987}).

\bibitem[{\citenamefont{Fischetti and Laux}(1996)}]{fischetti_band_1996}
\bibinfo{author}{\bibfnamefont{M.~V.} \bibnamefont{Fischetti}}
  \bibnamefont{and} \bibinfo{author}{\bibfnamefont{S.~E.} \bibnamefont{Laux}},
  \bibinfo{journal}{Journal of Applied Physics} \textbf{\bibinfo{volume}{80}},
  \bibinfo{pages}{2234} (\bibinfo{year}{1996}), ISSN \bibinfo{issn}{00218979}.

\bibitem[{\citenamefont{Sellier et~al.}(2007)\citenamefont{Sellier, Lansbergen,
  Caro, Rogge, Collaert, Ferain, Jurczak, and
  Biesemans}}]{sellier_subthreshold_2007}
\bibinfo{author}{\bibfnamefont{H.}~\bibnamefont{Sellier}},
  \bibinfo{author}{\bibfnamefont{G.~P.} \bibnamefont{Lansbergen}},
  \bibinfo{author}{\bibfnamefont{J.}~\bibnamefont{Caro}},
  \bibinfo{author}{\bibfnamefont{S.}~\bibnamefont{Rogge}},
  \bibinfo{author}{\bibfnamefont{N.}~\bibnamefont{Collaert}},
  \bibinfo{author}{\bibfnamefont{I.}~\bibnamefont{Ferain}},
  \bibinfo{author}{\bibfnamefont{M.}~\bibnamefont{Jurczak}}, \bibnamefont{and}
  \bibinfo{author}{\bibfnamefont{S.}~\bibnamefont{Biesemans}},
  \bibinfo{journal}{Applied Physics Letters} \textbf{\bibinfo{volume}{90}},
  \bibinfo{pages}{073502} (\bibinfo{year}{2007}), ISSN
  \bibinfo{issn}{00036951}.
  
\bibitem[{Note1()}]{Note1}
 \bibinfo{note}{Software is named for informational purposes only; it
  does not imply an endorsement or a recommendation by NIST.}



\end{thebibliography}

\def\url#1{}

\end{document}